**Combining multiplexed functional data to improve variant classification**

Jeffrey D. Calhoun[1], Moez Dawood[2,3,4], Charlie F. Rowlands[5], Shawn Fayer[6,7], Elizabeth J. Radford[8,9,10], Abbye E. McEwen[6,7,11], Malvika Tejura[6,7], Clare Turnbull[5,12], Amanda B. Spurdle[13,14], Lea M. Starita[6,7], Sujatha Jagannathan[15,16,*]

[1] Ken and Ruth Davee Department of Neurology, Northwestern Feinberg School of Medicine, Chicago, Illinois

[2] Human Genome Sequencing Center, Baylor College of Medicine, Houston, TX, USA

[3] Department of Molecular and Human Genetics, Baylor College of Medicine, Houston, TX, USA

[4] Medical Scientist Training Program, Baylor College of Medicine, Houston, TX, USA

[5] Division of Genetics and Epidemiology, The Institute of Cancer Research, London, UK

[6] Brotman Baty Institute for Precision Medicine, Seattle, WA, USA

[7] Department of Genome Sciences, University of Washington, Seattle, WA, USA.

[8] Wellcome Sanger Institute, Hinxton, CB10 1SA, UK.

[9] Department of Pediatrics, University of Cambridge, Level 8, Cambridge Biomedical Campus, Cambridge, CB2 0QQ, UK.

[10] Department of Pediatrics, Cambridge University Hospitals NHS Foundation Trust

[11] Department of Laboratory Medicine and Pathology, University of Washington, Seattle, USA

[12] The Royal Marsden NHS Foundation Trust, Fulham Road, London, UK

[13] Population Health Program, QIMR Berghofer Medical Research Institute, Herston, QLD, 4006, Australia

[14] Faculty of Medicine, The University of Queensland, Brisbane, QLD, 4006, Australia

[15] Department of Biochemistry and Molecular Genetics, University of Colorado Anschutz Medical Campus, Aurora, CO, USA.

[16] RNA Bioscience Initiative, University of Colorado Anschutz Medical Campus, Aurora, CO, USA.

* Corresponding author: sujatha.jagannathan@cuanschutz.edu

**Abstract:**

With the surge in the number of variants of uncertain significance (VUS) reported in ClinVar in recent years, there is an imperative to resolve VUS at scale. Multiplexed assays of variant effect (MAVEs), which allow the functional consequence of 100s to 1000s of genetic variants to be measured in a single experiment, are emerging as a powerful source of evidence which can be used in clinical gene variant classification. Increasingly, multiple published MAVEs are available for the same gene, sometimes measuring different aspects of variant impact. When multiple functional roles of a gene need to be considered, combining data from multiple MAVEs may provide a more comprehensive measure of the consequence of a genetic variant, which could impact variant classifications. Here, we provide guidance for combining such multiplexed functional data, incorporating a stepwise process from data curation and collection to model generation and validation. We demonstrate the potential and pitfalls of this approach by showing the integration of multiplexed functional data from five MAVEs for the gene *TP53*, two MAVEs for the gene *LDLR* and two MAVEs for *PTEN.* We also present a web applet that allows users to test various methods for combining score sets from multiple assays, calculate integrated functional scores for all variants, and assess whether combining data enables the application of stronger evidence for pathogenicity or benignity. By following these steps with appropriate guardrails, researchers can maximize the value of MAVEs, strengthen the functional evidence for clinical variant classification, and potentially uncover novel mechanisms of pathogenicity for clinically relevant genes.

## Background

Next generation sequencing is widely used clinically to provide a molecular diagnosis for genetic diseases. However, it is difficult to predict the functional impact, and therefore disease risk, associated with rare variants in Mendelian disease genes. We often lack sufficient information to classify these variants as pathogenic or benign and they become variants of uncertain significance (VUS). This is especially true for rare missense variants, >90% of which are currently classified as VUS in ClinVar (1). The high likelihood that a newly observed variant will be a VUS has made interpretation of genetic variants a substantial bottleneck in clinical genetics; absence of a causative pathogenic variant, and thereby a molecular diagnosis, limits effective counselling, management and treatment, restricting patient access to prophylactic, pharmaceutical and, looking to the future, potential gene therapy interventions. Variant interpretation is particularly challenging where other lines of evidence (*de novo* status, allele segregation, family history, etc.) are not available. Furthermore, interpretation of variants identified in individuals of non-European-like genetic ancestries is often limited by the poor diversity of our clinical and population databases, causing substantial inequity in diagnosis and treatment (2).

Experimentally derived variant effect data provide a valuable source of information to assist in variant classification. According to the American College of Medical Genetics and the Association for Molecular Pathology (ACMG-AMP) guidelines for interpretation of sequence variants, data from "well-established" functional assays can provide strong evidence towards pathogenicity and benignity for variant classification under the PS3 and BS3 criteria, respectively (3). However, reactive studies that model the function of individual variants cannot match the scale at which novel variants are discovered and are prohibitively laborious and expensive. Therefore, proactive, large-scale investigations of genetic variant effects are much needed.

As nucleic acid synthesis and sequencing has become more economical and accessible, a new generation of assays called multiplexed assays of variant effects (MAVEs) has evolved. MAVEs systematically assess the functional impact of thousands of variants, typically all SNVs or all amino acid substitutions, in a given gene or functional domain (4). While many MAVEs use pooled *in vitro* mutagenesis and introduction of an exogenous variant library into cells to assess function, approaches employing CRISPR/Cas9-based genome engineering such as Saturation Genome Editing (5-12), base editors (13-16), or prime editors (17, 18) are being increasingly employed to enable systematic editing of endogenous gene loci (19). This approach allows splice-modulating and other non-coding variant effects to be tested and enables variant effect assessment informed by the native sequence context.

Compared to traditional functional assays of individual variants, MAVEs offer substantial time and cost savings per-variant. In addition, the effect of an individual variant is interpreted in the context of the effect of all other possible variants in the target locus for the tested molecular phenotype. Hence, MAVEs can allow rigorous statistical analysis, including estimation of the assay sensitivity and specificity, and positive and negative predictive value, enabling appropriate weighting of the information within diagnostic variant classification frameworks (20, 21). For a MAVE measuring loss-of-function, setting assay thresholds using clinically classified benign and pathogenic missense variants (in addition to nonsense and synonymous assay control variants) ensures full dynamic range of scores as goalposts for defining functionality. Similarly, in the gain of function context, with a quorum of high-quality, known pathogenic variants with an analogous molecular phenotype, one can indicate the range for functionally deleterious variants in a MAVE. By assessing all variants in a gene, variant effect sub-groups can be identified, facilitating genotype-phenotype correlations which may be prognostically informative (22).

The last decade has seen an explosion of MAVEs measuring millions of variant effects that use different modalities to study variants in a variety of clinically important genes (23). In fact, there are now multiple MAVE datasets (hereafter referred to as multiplexed functional data) available for a number of genes (**Table 1**). The next decade is likely to see an increasing number of genes for which two or more multiplexed functional datasets are available. Thus far, guidance for using functional data in the clinic suggests using the dataset that is the best validated and ignoring or overriding the others (20, 21). However, combining experimental output data from two or more MAVEs may increase overall accuracy and clinical utility. Most simplistically, the impact of experimental noise may be mitigated by combining measurements across multiple assays. Furthermore, there are specific advantages to combining data generated using multiple phenotypic readouts or cell-types to enable a more comprehensive interpretation of variant effects.

| **Gene (CDS length)[REF(s)]** | **Genetic disorder** | **Missense variants in ClinVar** | **Missense variants in gnomAD (Z-score)** |
|---|---|---|---|
| *ASPA* (942 bp) (24) | Spongy degeneration of central nervous system | B/LB=7; P/LP=73; VUS=62 | 315 (1.25) |
| *BRCA1* (5592 bp) (6) | Hereditary breast ovarian cancer syndrome | B/LB=384; P/LP=266; VUS=1,552 | 1948 (2.34) |
| *BRCA2* (10257 bp) (25, 26) | Hereditary breast ovarian cancer syndrome | B/LB=369; P/LP=96; VUS=3,596 | 3781 (1.02) |
| *CARD11* (3465 bp) (8) | Severe combined immunodeficiency due to *CARD11* deficiency | B/LB=80; P/LP=17; VUS=340 | 1180 (3.73) |
| *CHEK2* (1632 bp) (27, 28) | Hereditary cancer-predisposing syndrome | B/LB=9; P/LP=12; VUS=1,651 | 486 (-0.02) |
| *F9* (1386 bp) (29) | Hereditary factor IX deficiency disease | B/LB=25; P/LP=186; VUS=66 | 241 (N/A*) |
| *GCK* (1398 bp) (30, 31) | Monogenic diabetes | B/LB=12; P/LP=335; VUS=217 | 451 (2.79) |
| *LDLR* (2583 bp) (32) | Familial Hypercholesterolemia | B/LB=44; P/LP=739; VUS=777 | 1484 (0.61) |
| *NUDT15* (495 bp) (33-35) | Poor metabolism of thiopurines | B/LB=1; P/LP=0; VUS=6 | 214 (-0.59) |
| *PTEN* (1212 bp) (36, 37) | Hereditary breast ovarian cancer syndrome | B/LB=26; P/LP=277; VUS=742 | 200 (4.12) |
| *SOD1* (465 bp) (38) | Amyotrophic lateral sclerosis | B/LB=0; P/LP=112; VUS=51 | 150 (1.22) |
| *TARDBP* (1245 bp) (39, 40) | Amyotrophic lateral sclerosis type 10 | B/LB=9; P/LP=28; VUS=87 | 202 (5.32) |
| *TP53* (1182 bp) (41-43) | Li-Fraumeni syndrome | B/LB=144; P/LP=336; VUS=609 | 464 (1.15) |

**Table 1: Exemplar genes with two or more multiplex functional datasets currently available.** MAVEregistry (https://registry.varianteffect.org/) was accessed on 8 Oct 2024 to generate this list. Additional genes were manually curated. The ClinVar database was accessed 22 Jan 2025 to obtain the number of variants classified as VUS. gnomAD v4.1 data was used to collect the number of missense variants per gene and the missense Z-score (a functional constraint metric (44) based on observed versus expected number of missense variants in a gene, where high positive scores indicate intolerance to missense variation.) *No gene constraint metrics are currently available for *F9* in gnomAD. Genes with multiplexed functional data from a single assay with multiple measurement sets include, but are not limited to: *CBS*, *DDX3X*, *RAD51C*, *BAP1*, *UBB*/*UBC*, *VHL*, *TP53*, and *BRCA1*.

While integrated analyses have been published for specific genes (45-47), systematic guidance for when and how to combine MAVEs has been lacking. Here, we discuss two scenarios in which combining variant effect data across MAVEs may be beneficial and, conversely, describe scenarios where combining such multiplexed functional data would not be recommended. We provide a set of recommendations and best practices for selecting and combining multiplexed functional data to improve their clinical utility in variant classification. Finally, we provide three practical examples where different methods are used to integrate multiplexed functional data from five MAVEs for the gene *TP53*, two MAVEs for *LDLR,* and two MAVEs for *PTEN*.

**Glossary**

B/LB = benign or likely benign
CDS = coding sequence
DMS = deep mutational scan
DN = dominant negative
GOF = gain-of-function
Gene-Disease Dyad = Specific pairing of a gene to a specific clinical phenotype
Indel = insertion or deletion
LOF = loss-of-function
MAVE = multiplexed assay of variant effect
Molecular Phenotype = Phenotypic outcome measured by the MAVE
NGS = next generation sequencing
NPV = negative predictive value
OddsPath = odds of pathogenicity
P/LP = pathogenic or likely pathogenic
PPV = positive predictive value
PTC = premature termination codon
SGE = saturation genome editing
SNV = single nucleotide variants
VAMP-seq = variant abundance by massively parallel sequencing
VUS = variant of uncertain significance

## Methods

*Development of recommendations.* The recommendations presented here grew out of discussions among members of the Clinical Variant Interpretation (CVI) Working Group within the Atlas of Variant Effects (AVE) Alliance, a group of researchers and clinicians generating or applying multiplexed functional data across the world. An initial set of recommendations developed by a subgroup (Calhoun, Dawood, Fayer, Radford, Rowlands, and Jagannathan) was presented at multiple CVI working group meetings to further refine a set of specific recommendations. Subsequent iterations of the recommendations involved conversations among expert clinicians, leaders in ClinGen and MAVE experts (Starita, McEwen, Roth, Turnbull, and Spurdle) before agreement on the final content of the recommendations was reached.

*Curation of clinical truth sets for TP53, LDLR, and PTEN.* A major concern when working with well-studied genes such as *TP53* is the possibility of circularity within truth sets, where variants classified using a certain functional assay are used as "ground truth" to calibrate and validate the same or highly similar functional datasets. To avoid this circularity and maintain evidence independence, we used a truth set curated to only contain *TP53* variants which could be classified in the absence of functional and computational data (48). In the original study, *TP53* variants were assigned likelihood ratios (LRs) based on clinical data, functional assays, and bioinformatic tools. These LRs were combined and used to calculate a posterior probability, which was used to assign variant classifications. We modified this dataset by removing all LRs from functional datasets and bioinformatic tools. We then applied a naïve prior probability of pathogenicity of 0.1 and recalculated the posterior probability using only the remaining clinical LRs. Variants were reclassified based on the following posterior probability thresholds: Pathogenic (>0.99), Likely pathogenic (0.90-0.99), Likely Benign (0.001-0.010), Benign (< 0.001) (49). The new classifications were then used as an updated truth set for downstream analyses. The *LDLR* truth set is identical to that used in the LDLR MAVE publication and consists of benign or likely benign (B/LB) and pathogenic or likely pathogenic (P/LP) missense variants retrieved from ClinVar (32). The *PTEN* truth set was similarly curated by retrieving B/LB and P/LP missense variants from ClinVar (accessed 11 Jun 2025).

*Development of web Shiny application and demonstration of recommendations using TP53, LDLR, and PTEN data.* A web Shiny application was developed to demonstrate multiplexed functional dataset integration herein and facilitate sharing with the broader scientific community. Input datasets were curated where each row comprises a different variant and each column is the average functional score for one input MAVE (41, 42). For the purposes of this demonstration, only variants scored by all assays were included. Clinical truth sets were supplemented with different fractions of nonsense and synonymous assay controls ("proxy" truth set) depending on the size of the clinical truth set. For *TP53*, which has a robust clinical truth set, no assay controls were included in model training. For *LDLR*, the clinical truth set was supplemented with a random selection of 50% of available synonymous and premature termination codon variants for model training. For *PTEN,* the entire set of assay controls were included in training after filtering out PTCs in the last exon and the last 50 nt of the penultimate exon, which are known to escape nonsense-mediated RNA decay and may not induce LOF (50). Integration methods used in the illustrative examples include: (1) principal component analysis (PCA); (2) unsupervised k-means clustering; (3) supervised naive Bayesian classifier; and (4) a random forest (n=5 trees) supervised classifier. (50) Integration methods used in the illustrative examples include: (1) principal component analysis (PCA); (2) unsupervised k-means clustering; (3) supervised naive Bayesian classifier; and (4) a random forest (n=5-10 trees) supervised classifier. For *TP53*, supervised learning methods used a randomized 75%/25% train/test split, while 70%/30% and 50%/50% train/test splits were used for *LDLR* and *PTEN*, respectively. For k-means clustering, the output column is a categorical variable for cluster; for the other methods, the output column is a single integrated score. Model metrics are computed using a single threshold determined either by ROC analysis to best distinguish B/LB variants from P/LP variants, or set manually by inspecting the separation of B/LB and P/LP variants in a histogram of functional scores. Evidence strength was determined using the odds of pathogenicity (OddsPath) framework described by Brnich et al. (20), which uses Bayesian statistics to convert functional assay performance into likelihood ratios for clinical variant classification within the ACMG/AMP framework. All the model metrics, threshold values used by various classifiers, the variant-level integrated scores assigned by different methods are downloadable as tables through the Shiny application.

## Points of Consideration and Recommendations

We will focus on two scenarios in which combining multiplexed functional data may be beneficial.

1. **One gene, different MAVEs**: A genetic condition may be caused by more than one genetic mechanism, such as loss of function (LOF) and gain of dominant negative (DN) function. However, some MAVEs may only be capable of measuring either the LOF mechanism or the DN, but not both. Even within the LOF mechanism, many MAVEs measure only one aspect of protein function. For example, variant abundance by massively parallel sequencing (VAMP-seq) assays measure protein stability but are blind to enzymatic activity (36). Thus, despite the excellent positive predictive value (PPV) to identify a pathogenic variant, the ability to confidently determine a variant as benign (negative predictive value) of such functional assays is impaired. In such cases, combining multiplexed functional data measuring multiple aspects of protein function to generate a single integrated score may improve the accuracy of functional predictions for that gene.

2. **One MAVE, many measurement sets**: Within a MAVE, the dynamic range can be extended by collecting data at multiple points of time to provide quantitative information on a phenotype that may improve variant classification (9, 10, 12). A similar argument could be made for MAVEs that measure the same molecular phenotype but through different assays, in different model systems and cell lines, or at different concentrations of a drug or other treatment that may distinguish different degrees of gene function. By quantifying variant effects over time or across different experimental conditions, putative hypomorphic variants that cause partial functional deficits could be potentially distinguished from complete LOF variants.

**When not to combine multiplexed functional data:** There are scenarios where it may not be desirable to combine multiplexed functional data. For example, distinct disorders can arise through different mechanisms of genetic variants within the same gene (i.e. distinct gene-disease dyads (51)). In the case of *CARD11* gene, gain-of-function, loss-of-function, and dominant negative variants result in distinct genetic disorders (52). As a measure of functional variant effect for use in clinical variant classification, separate scores and/or thresholds may be required for each gene-disease dyad. It is also important to consider technical aspects of MAVE design which determine the type of variants that can be assessed. Some MAVEs are based on a transgenic cDNA platform, whereas others that use genome editing of an endogenous allele are "splice-aware". Care must be taken when integrating data to ensure the results from a cDNA assay do not override a genome editing assay for variants that impact splicing.

**Considerations**

Hypothetically, a single MAVE would distinguish every P/LP variant from every B/LB variant for a gene-disease dyad, presuming there is a single mechanism of pathogenicity. In practice, there is usually overlap between functional scores for some known functionally normal and functionally abnormal variants (**Figure 1A-B**). Despite this overlap, it is usually possible, at least for loss of function (LOF) assays, to set thresholds which adequately distinguish the majority of synonymous variants from nonsense variants, as well as variants previously classified clinically as B/LB from those classified as P/LP. There are several ways to assess assay performance at this stage: (1) the dynamic range to separate functionally normal and abnormal variants, and (2) computing standard classification metrics (sensitivity, specificity, PPV, NPV, etc. shown in **Figure 1C**). These metrics establish a baseline for each individual dataset, which is critical to later investigate how combining multiplexed functional data impacts these metrics.

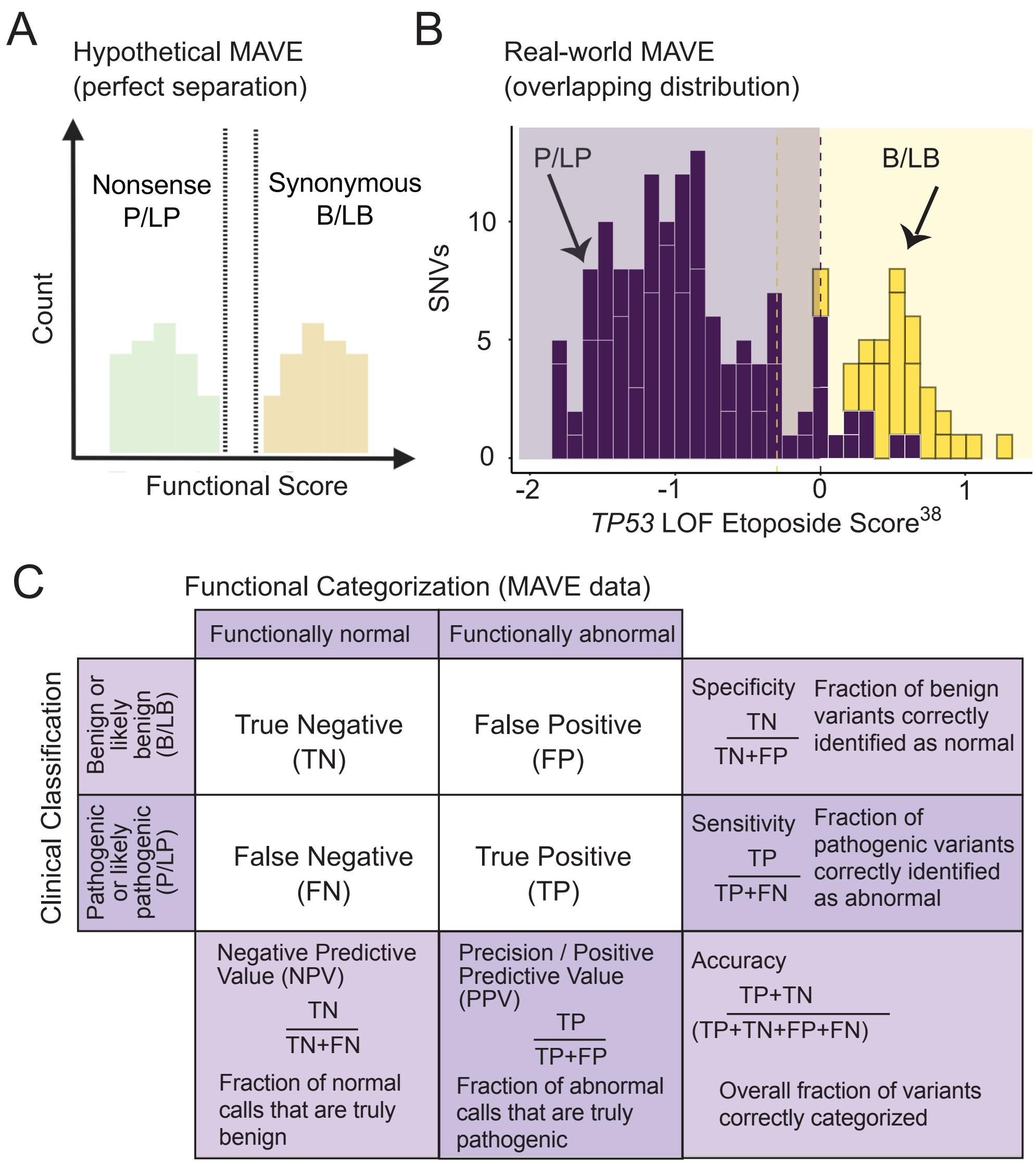

**Figure 1: Assessment of assay performance prior to combining multiplexed functional data.** (A) Functional score distributions for a hypothetical perfect multiplexed functional assay for assumed functionally normal (synonymous) & B/LB variants and assumed functionally abnormal (premature termination codon, i.e. PTC) & P/LP variants. (B) An example of real-world multiplexed functional data illustrating overlap between the P/LP and B/LB variants scored in the assay (41). (C) Critical metrics for assessing MAVE performance.

Clinical genetic diagnostics currently uses a specific framework outlined in 2015 by the ACMG/AMP in order to robustly ascribe pathogenicity or benignity to individual variants.(3) In this framework, different sources of evidence are applied at varying strengths and combined to reach a final classification. Functional scores from a MAVE can currently be incorporated as one source of evidence via the PS3 and BS3 criteria (towards pathogenicity and benignity, respectively) in the ACMG/AMP framework. The strength of applicable evidence is determined using the Bayesian framework described by Brnich et al (20) and depends on assay performance metrics calculated using a clinically validated truth set. The Brnich framework calculates likelihood ratios (termed "OddsPath") by comparing the prior probability of pathogenicity in the truth set (proportion of P/LP variants in the truth set) with the posterior probability of pathogenicity among variants falling above or below a defined threshold for functionality (proportion of P/LP variants among those categorized as functionally abnormal or normal, respectively). The prior and posterior probabilities are then used to calculate assay-specific likelihood ratios (OddsPath) towards pathogenicity and benignity (20). These likelihood ratios can be calculated for both individual multiplexed functional assays and integrated scores to assess whether data integration improves evidence strength and enables application of stronger evidence categories within the functional evidence range.

To combine multiplexed functional data to generate an integrated score that can provide stronger evidence to improve clinical variant classification, some basic criteria must be met:

1. **The relationship between what the assay measures and the human phenotype are the same for both assays.** To combine different multiplexed functional data for a particular gene-disease dyad, it is important to assess whether the molecular phenotypes of each MAVE being combined are relevant to the disease and overall gene-disease dyad, even if the precise molecular mechanisms are not fully understood. For example, a cell growth-based *CARD11* MAVE measured a LOF phenotype linked to immunodeficiency disorders such as severe combined immunodeficiency (SCID) (8). Conversely, measuring the response of cells after treatment with the BTK inhibitor ibrutinib enabled the same cell-based system to assess a GOF phenotype linked to the lymphoproliferative disorder BENTA (8). Therefore, the assays are measuring variant effects relevant for different clinical phenotypes and likely distinct molecular mechanisms and should be assessed independently and not combined.

2. **The multiplexed functional data being combined provide complementary information**. Combining datasets generated by the same assay performed in two different laboratories may yield some benefit in providing a better estimate of the true functional effect across more measurements, but the data are likely to be highly correlated and additional information from the integrated score may be limited. Conversely, two or more assays which measure different aspects of molecular function associated with the same clinical phenotype may be more likely to improve variant classification outcomes when combined by providing a more complete picture of overall protein function, even when the assays show some correlation. For instance, independent assays have been developed quantifying the impact of *PTEN* variants on both the abundance (36) and enzymatic activity (37) of the PTEN protein. Similarly, quantification of both protein abundance and thiopurine toxicity have been utilized as independent readouts to investigate the impact of variants in *NUDT15* (35). Work by Cagiada et al. has shown that around half of variants leading to loss-of-function in these assays additionally lead to reduced protein abundance (33). In combining multiplexed functional data for these genes, abundance might therefore serve as a complementary readout for a subset of loss-of-function variants (33).
3. **A sufficient number of truth set variants are available for assay calibration** (20). Construction of truth sets for combined multiplexed datasets may not be feasible where the loci targeted by each dataset are distinct, e.g. when constituent MAVEs target specific functional domains. In the absence of a robust clinical truth set, it may be possible to calibrate based on synonymous (presumed benign) and nonsense (presumed pathogenic) variants if the disease mechanism is loss of gene function. However, the strength of evidence that can be applied in the ACMG/AMP framework is likely to be limited without a sufficiently large clinically relevant truth set.

We acknowledge that criteria 1 and 2 are not always easy to assess. In that case, we recommend being guided by the ability of the combined score to accurately categorize truth set variants as normal or abnormal compared to the individual MAVEs.

**<u>Practical considerations for integrating multiplexed functional data to generate an integrated score</u>**

A general workflow for integrating multiplexed functional data comprises: i) Collecting, harmonizing, and merging multiplexed functional data; ii) Collecting and harmonizing truth set;

iii) Choosing an appropriate method to generate an integrated score; iv) Setting thresholds for functionally normal and abnormal variants; and v) Final assessment of whether the integrated score has improved utility for variant categorization (**Figure 2A**).

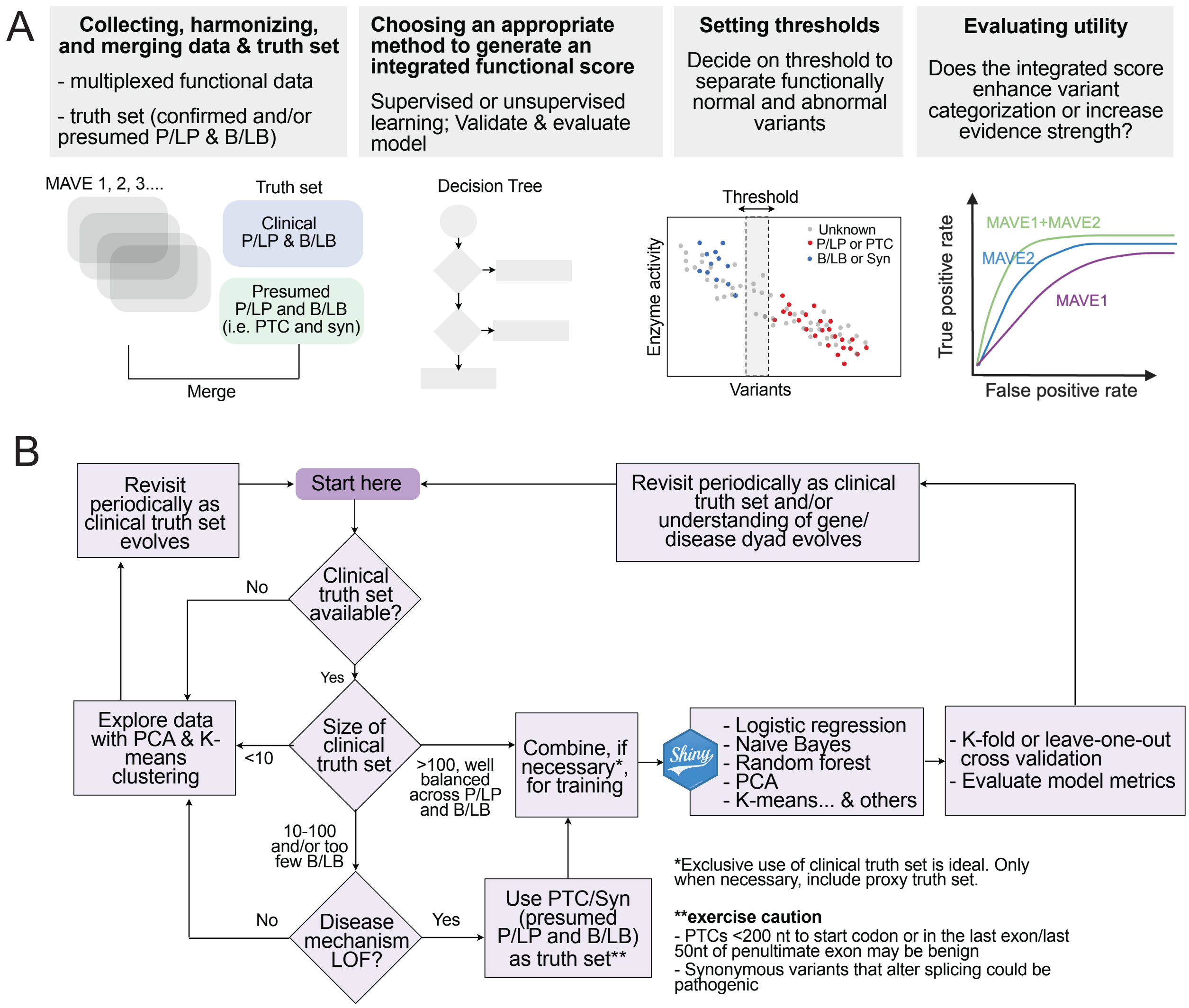

**Figure 2: Suggested pipeline and decision tree to generate and evaluate integrated functional score. A)** A method for combining data will involve merging the datasets from different MAVEs and the truth set. A final score for each variant will be generated using an appropriate supervised or unsupervised method. Threshold values to separate functionally normal and abnormal variants will be set based on truth set. After thresholding, the integrated score must be assessed for improved variant resolution compared to the multiplexed functional data from individual MAVEs based on how well the different scores distinguish B/LB and P/LP variants in the truth set. **B)** Decision tree approach to choosing an appropriate method to combine multiplexed functional data.

*i)* Collecting, harmonizing, and merging multiplexed functional data

Prior to merging datasets, it is important to ensure that each dataset uses standardized variant nomenclature identifiers like the Human Genome Variation Society (HGVS) nomenclature and that each dataset references the same canonical gene transcript(s) prior to merging. Tools such as Transvar can be used to harmonize data, ensuring that variants are matched to a common transcript (53). Next, decide if only variants evaluated across all individual MAVEs will be retained or if missing data will be allowed. Relatedly, decide if any variants should be excluded from specific MAVEs, such as variants predicted to impact splicing from cDNA-based assays. See **Section iii** for more guidance on both points. Lastly, one may wish to transform the data prior to modeling. For example, standardization, or scaling each input dataset to a common scale (typically between 0 and 1), is commonly employed prior to supervised learning methods. While linear rescaling is most commonly applied, depending on the specific context of the MAVE, a non-linear rescaling method may be most appropriate. Prior to standardization, it can be useful to perform normalization (via log or other transformation), in particular if a dataset is not normally distributed.

*ii)* Collecting and harmonizing truth set

Truth set variants will be used both in model training (for supervised learning) and to evaluate how well the final integrated score performs in categorizing variants of known pathogenicity and benignity as abnormal and normal, respectively. It is highly recommended, when feasible, to collect truth set variants from an appropriate clinical database such as ClinVar (54) or LOVD (55), keeping in mind that the classification of some of these variants may have used one or more of the functional assays being evaluated, introducing circularity. Variants classified exclusively using bioinformatics tools or population frequency data should also be used with caution to maintain evidence independence, if those same tools or data sources will be used in the final variant classification process alongside the functional evidence. We recommend manually curating variants to only retain those that were not classified using functional evidence (as we do for the *TP53* example). Where this is not possible, one could obtain a historic truth set from an archived version of a clinical database from a time prior to the release of the functional data, although doing so will likely reduce the size as well as the quality of the truth set considerably.

When clinically validated truth sets of sufficient size are unavailable, as is often the case, synonymous (presumed benign) and nonsense (presumed pathogenic) variants included in a multiplexed functional assay are used as a proxy truth set, if the disease mechanism is loss of function (6, 37, 45). There are key caveats to using such a proxy truth set, including the loss of dynamic range in ascertaining B/LB or P/LP variants that are partially functional as well as misannotation of certain nonsense variants as P/LP when they are truly benign (56) or synonymous variants as B/LB when they may induce missplicing of a transcript (57). Pre-emptive filtering of synonymous variants predicted to impact splicing and nonsense variants that occur in the last exon or within 50 nucleotides of the last exon-exon junction (58, 59) from proxy truth sets is recommended in such cases. Finally, while the OddsPath framework can be applied to either clinically ascertained truth sets or proxy truth sets (or a combination), robust evidence strength assignment and clinical utility require clinically validated truth sets.

*iii)* Choosing an appropriate method to generate an integrated functional score

In choosing a method for combining multiplexed functional data and generate and integrated functional score, a user must carefully consider their end goal. If interpretability of a model is an important consideration, statistical methods such as logistic regression, which allows quantification of the weighting of covariates in the combined model is recommended. If performance is the primary concern, a more complex but less-interpretable supervised model such as naive Bayes or random forest classifier may be used. If the goal is to explore patterns without labeled data, unsupervised methods like k-means clustering or PCA would be appropriate.

Beyond the primary goal, there are practical considerations that may influence the method of choice. When two MAVEs measure different features of a protein - for instance, where one reports on protein stability while another reports on protein function, it is important to carefully consider the gene function and disease mechanism to assess if these two signals might be conflicting before choosing an integration methodology. For example, genomic assays are more relevant to the classification of splice-impacting variants compared to cDNA-based assays. In such cases, supervised methods such as logistic regression, naive Bayes, or random forest automatically weigh features (in this case different multiplexed functional datasets) based on their ability to classify clinical truth sets. On the other hand, methods such as PCA will need to be manually weighted based on individual MAVE performance metrics (e.g., sensitivity, specificity) or OddsPath values calculated on held-out validation data. However, the choice of weights must be carefully considered and justified to avoid introducing user bias.

An alternate approach is to remove specific variants from certain multiplexed function datasets to ensure that conflicting signals are resolved prior to combining the data. For example, computational prediction tools (e.g., SpliceAI (60), Pangolin (61), MaxEntScan (62)) can be used to flag and exclude splice-affecting variants from cDNA-based assays prior to integration. However, any such exclusions must be documented transparently. A related consideration is whether an end user wants to only use complete cases (i.e. data is available for the variant in all MAVEs) or would like to use all data irrespective of some missing values. Methods like random forest classifiers handle missing values through median imputation within each class during training. However, this approach may not be appropriate for clinical decision-making. We recommend using only complete cases (variants scored by all MAVEs) for clinical classification, while exploratory analyses may include all available data.

Finally, the size of the truth set will also influence which models are appropriate to use (**Figure 2B**). For genes with small truth sets (10-100 variants), complex supervised learning methods, such as the random forest method, are not recommended as such methods often require a training/test split and too few variants will remain in the test set to adequately assess categorization. Instead, statistical methods or unsupervised learning methods are recommended in these cases. For genes with larger truth sets, complex supervised learning methods may be explored in addition to statistical methods and unsupervised learning.

*Supervised methods*

Supervised approaches, including both statistical and machine learning methods, use labeled data to train a model to predict a specific outcome, in this case normal versus abnormal categorization of B/LB vs. P/LP truth set variants. These approaches can be powerful but require a truth set of sufficient size (which varies by method) to train on. The training set will be used to train the machine learning classifier, while the held-out test set is used to evaluate the accuracy of the classifier on unseen data that was not used in training. K-fold or leave-one-out cross validation is recommended to avoid overfitting.

Logistic regression is a more readily interpretable supervised approach that can categorize variants as normal or abnormal in function. Logistic regression calculates a coefficient for each variable that represents the importance of that specific feature or dataset to the final categorization. Logistic regression can perform well even for smaller datasets and with lesser computational resources. One key drawback of logistic regression is that it cannot handle missing data points, which would be an issue on combining multiple MAVEs that assess non-overlapping sets of variants.

Machine learning approaches such as naive Bayes classifiers and random forest methods, while not readily transparent to interpret, can perform very well on large datasets. Such approaches can be advantageous where there is a non-linear relationship between the features and the outcome, as well as interactions across multiplexed functional datasets. For example, random forest classifiers can handle complex relationships between the different datasets whereas naive Bayes approaches assume that the different features are completely independent, which may not be appropriate in some scenarios.

In one published example, a naive Bayesian classifier was used to classify variants based on combining multiplexed functional data for the gene *TP53* to generate a single functional prediction (41, 42, 45). None of the constituent MAVEs perfectly discriminated known pathogenic from benign variants, with the greatest applicable evidence strength for any constituent assay being PS3_moderate. By contrast, the combined classifier approach achieved an OddsPath of 30.3 for functionally abnormal variants and 0.054 for functionally normal variants. This enabled application of strong evidence towards pathogenicity (PS3) and moderate evidence towards benignity (BS3) for the combined model. Combining predictions from this classifier with other available data enabled reclassification of 69% of *TP53* VUS (45).

*Unsupervised approaches*

Unsupervised methods use unlabeled data to find patterns without external training data. When there is no truth set available (i.e. no labeled data to train on) or the size of the truth set is insufficient for supervised machine learning approaches, unsupervised approaches such as PCA and k-means clustering could still identify patterns in the data that might inform the consequence of a variant either in terms of function or in terms of pathogenicity. Unsupervised clustering algorithms such as t-SNE and UMAP, which have been widely adopted for the analysis of single cell RNA sequencing data (scRNA-seq) (63), enable visualizations of cell type clusters based on shared gene expression patterns. When applied to multiplexed functional data, these algorithms are not skewed by external factors such as training datasets derived from ClinVar which can be biased by ascertainment and ancestry (1, 64, 65). Most unsupervised methods are relatively immune to overfitting and can work well in cases where truth sets are unavailable or limited. Another advantage of unsupervised methods is that they allow the full truth set to be used (i.e. without needing to split into training and test sets) to validate the model and determine strength of evidence for clinical variant classification.

Finally, unsupervised approaches like Gaussian mixture modeling are better at capturing complex or high-dimensional relationships between multiplexed functional datasets such as time course data or an assay that can read out variable degrees of function to identify subgroups of variant classes instead of a binary classification. For example, in multiple studies, employment of Gaussian mixture modeling of multiplexed functional data collected across a time course has identified variants that appear to exert a milder effect *in vitro* (growing more slowly compared to other variants) than full loss-of-function variants (9, 10). Radford et al. demonstrated separate populations of fast- and slow-depleting *DDX3X* variants, where slow-depleting variants were hypothesized to exert milder defect on protein function (10). Confirmation of whether such observations are relevant to patient phenotypes requires large well-phenotyped clinical cohorts, which are lacking for most rare-disease genes.

*Model performance metrics*

The performance of a specific model can be evaluated by comparing the metrics outlined in **Figure 1C**, such as specificity and sensitivity, relative to the individual input MAVEs. Additionally, one must be cautious of the potential for overfitting when combining multiplexed functional data to maximize apparent performance in clinical variant discrimination. Exploring multiple combination strategies and selecting the one that performs best on the test set risks inflating apparent performance, which may not generalize to other variants. To alleviate this issue, we recommend evaluating variant classification via K-fold or leave-one-out cross validation. While the above framework provides a starting point for a user to consider the different approaches to combine MAVEs, users should adapt it based on assumptions made by each method, their relative strengths and weaknesses, as well as the intended application of the results. In the end, no single approach will be suitable for every use-case, and we recommend each user to be guided by their own domain expertise on gene function, the specific MAVEs being combined, and the gene-disease dyad (see the decision tree in **Figure 2B** to guide method selection).

When combining multiplexed functional data, one may observe improvements in some performance metrics while others decline. In such cases, OddsPath as well as clinical context should guide the selection of the right model. For example, when a clinical diagnosis of a condition has already been made (say, for cystic fibrosis), prioritizing sensitivity (successfully identifying true pathogenic variants) over specificity (avoiding mis-classification of truly benign variants as pathogenic) may be appropriate, if it would give a child access to specific treatments. Similarly, when PPV and NPV move in opposite directions, it becomes important to consider whether the clinical context calls for better pathogenic calls or better benign calls. A key variable that influences all these metrics is the threshold value(s) set to call a variant abnormal or normal with respect to function (see Section iv for more guidance on threshold setting). Ultimately, OddsPath values can serve as tie-breaker criteria in these cases because they translate directly to clinically applicable evidence strength categories (PS3/BS3) and provide a standardized framework for evaluating assay performance in both pathogenic and benign directions.

*iv) Setting thresholds to separate functionally normal and abnormal variants*

In general, the goal for setting thresholds is to achieve high confidence and the strongest possible evidence strength for or against pathogenicity. While single threshold methods are simpler to implement and allow categorization of all variants, they can lack nuance and have increased potential for misclassification near the threshold. Double threshold methods that use separate thresholds for benign and pathogenic categorization require more user discretion and result in more "indeterminate" calls but can minimize misclassifications if implemented appropriately. Clinical context and the potential consequence of misclassification should be carefully considered in choosing the appropriate thresholding strategy.

In many cases, the distributions of scores for assay control variants (such as nonsense and synonymous variants) are used to set thresholds. However, this approach can be problematic for the same reasons as using such variants as the truth set is problematic (described in Section ii). To avoid overreliance on assay control variants, we recommend using the score distribution of clinically validated P/LP and B/LB variants wherever possible to select appropriate thresholds.

There are multiple schools of thought on how to best set thresholds for separating functionally normal from functionally abnormal variants. In one approach, percentiles are used to determine thresholds. In Boyle et al (66), the 5$^{th}$ percentile of functional scores for synonymous *CYP2C19* variants was used as a threshold distinguishing between “WT-like” and “decreased” scores. In other studies, statistical modeling of variant distributions has been used. For instance, Nagy et al (67) modeled scores of missense variants in a control condition as a normal distribution and a threshold was selected based on the bottom first percentile of the distribution. Alternatively, receiver operating characteristic (ROC) curve analysis can be used to identify thresholds that optimize the balance between sensitivity and specificity based on known pathogenic and benign variant sets. While the examples given here are for individual multiplexed functional datasets, the same principles apply to an integrated dataset. Similar methods can be used to set thresholds for the integrated score which distinguish functionally normal from functionally abnormal variants. Regardless of the threshold-setting approach used, it is important to validate threshold performance using independent, clinically validated variant truth sets.

*v) Evaluating the utility of the integrated evidence strength assignment*

The performance of an integrated score or classification relative to each individual MAVE should be assessed to determine if the integrated score or classification has improved relative to the individual assays (**Figure 2**). This includes evaluating whether the difference between the dynamic ranges of functionally normal and abnormal variants increases when using the integrated score relative to the best-performing individual MAVE. If the integrated score does not lead to a clear improvement over the best-performing input MAVE, we recommend instead using the score for the best-performing individual MAVE when determining which functional score to translate into functional evidence for clinical variant classification.

It is important to emphasize that combining multiplexed functional data will not necessarily lead to an improved clinical utility, such as an increased number of variants reaching higher evidence strength, for every gene-disease dyad. In such cases, there may still be a use case for combining multiplexed functional data in a basic research context, as it may yield better understanding of protein structure-function relationships and/or identify unexpected variant clusters which yield novel insight into biology. For example, a gene with multiplexed functional data from multiple MAVEs spanning function, protein abundance, subcellular localization, and protein-protein interaction might reveal distinct functional classes of variants, such as loss of function versus gain of function or dominant negative. Functional categorization information could then feedback to clinical practice through improved understanding of disease mechanism.

## Results

To illustrate some of the concepts above, we present three brief examples of combining multiplexed functional data with different methodologies for the genes *TP53, LDLR,* and *PTEN*. We generated these examples using the shiny application we created for the purpose of easily evaluating methods to combine such data and evaluate the utility of an integrated score (**Figure 3,** and **Figure S1**). For these analyses, we focused on determining whether integration enhanced clinical utility by achieving higher evidence strength categories (as determined by OddsPath calculations), restricting analysis to variants scored across all contributing assays.

For *TP53*, we used five MAVEs from three publications (41-43). These datasets were deliberately selected for combination as they span multiple different cell types (A549, MOLM13, and HCT116) and utilize both positive and negative selection using compounds nutlin-3 or etoposide depending on the expression of functional *TP53* (41, 42). Further, three of these experiments were designed to screen for LOF variants while another two were designed for DN variants. Pathogenic variants in *TP53* are known to act by both LOF and DN mechanisms. We analyzed these datasets with a clinical truth set of 281 missense variants (223 B/LB and 58 P/LP) carefully selected to avoid classifications that relied on functional data (see Methods above for detailed description). From this truth set, 201 variants (143 B/LB and 58 P/LP) were scored in all five *TP53* MAVEs (**Figure 3A**). As this represents a well-balanced and appropriately sized truth set, we used 70% of these variants to train the model, with negligible inclusion of assay controls (1% of synonymous and PTC variants) in the training data.

To explore these data, we plotted missense B/LB and P/LP variants in principal component space based on their scores in the five individual MAVEs, as well as based on K-means clusters (**Figure 3B-C**). These data form distinct clusters, suggesting underlying patterns that distinguish B/LB and P/LP variants. We then employed both unsupervised and supervised learning methods to generate a single score representing the integration of all five multiplexed functional datasets, and evaluated both individual and integrated scores to assess performance (**Figure 3D;** see **Tables S1-S5** for classifier outputs, threshold values, and comprehensive model metrics). Individual MAVEs showed variable performance, with sensitivity ranging from 0.897 to

A

| *TP53* Truth Set | P/LP | B/LB | PTC | Syn | Ratio |
|---|---|---|---|---|---|
| | 58 | 143 | 394 | 378 | 0.7/0.01 |

B

Variants grouped by class

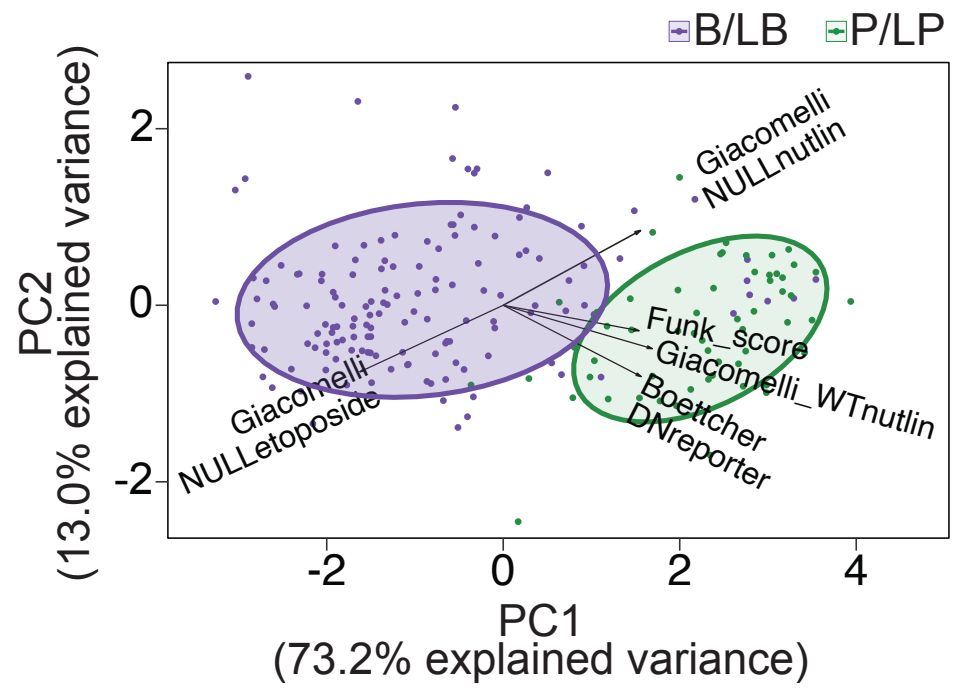

C

Variants grouped by k-means clusters (n = 4)

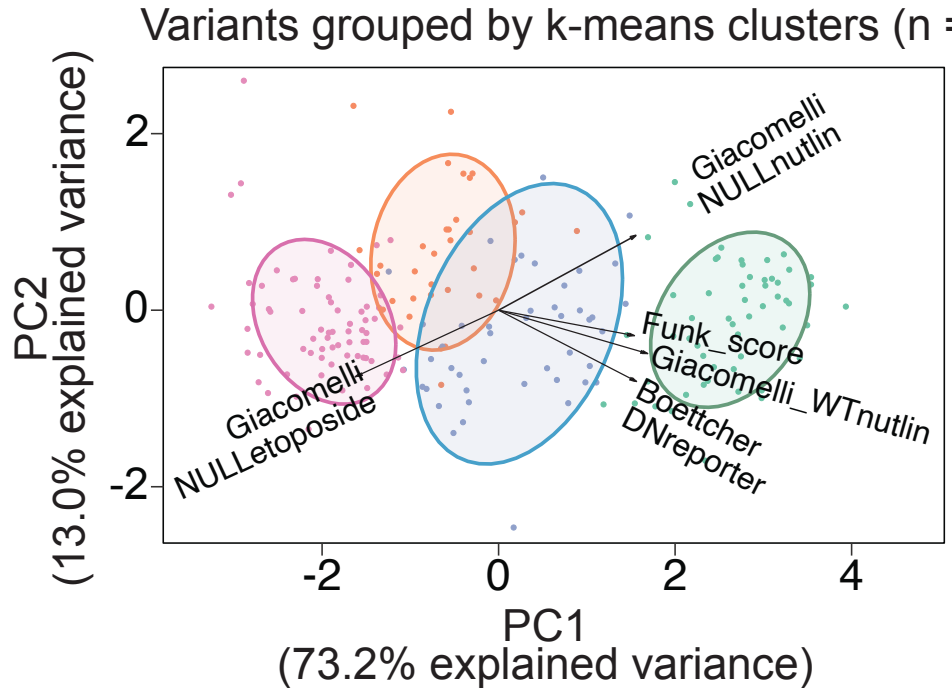

D

| *TP53* | Sens. | Spec. | PPV | NPV | AUC |
|---|---|---|---|---|---|
| MAVE1: Etoposide | 0.897 | 0.783 | 0.627 | 0.949 | 0.892 |
| MAVE2: Null_nutlin | 0.931 | 0.629 | 0.505 | 0.957 | 0.825 |
| MAVE3: WT_nutlin | 0.931 | 0.867 | 0.740 | 0.969 | 0.942 |
| MAVE4: DN_reporter | 0.914 | 0.944 | 0.869 | 0.964 | 0.960 |
| MAVE5: Funk | 0.931 | 0.811 | 0.667 | 0.967 | 0.905 |
| PCA | 0.948 | 0.881 | 0.764 | 0.977 | 0.948 |
| Naive Bayes (train) | 1.000 | 0.906 | 0.773 | 1.000 | 0.974 |
| Naive Bayes (test) | 0.958 | 0.865 | 0.821 | 0.970 | 0.944 |
| Random forest (train) | 1.000 | 0.981 | 0.944 | 1.000 | 1.000 |
| Random forest (test) | 0.917 | 0.892 | 0.846 | 0.943 | 0.954 |

E

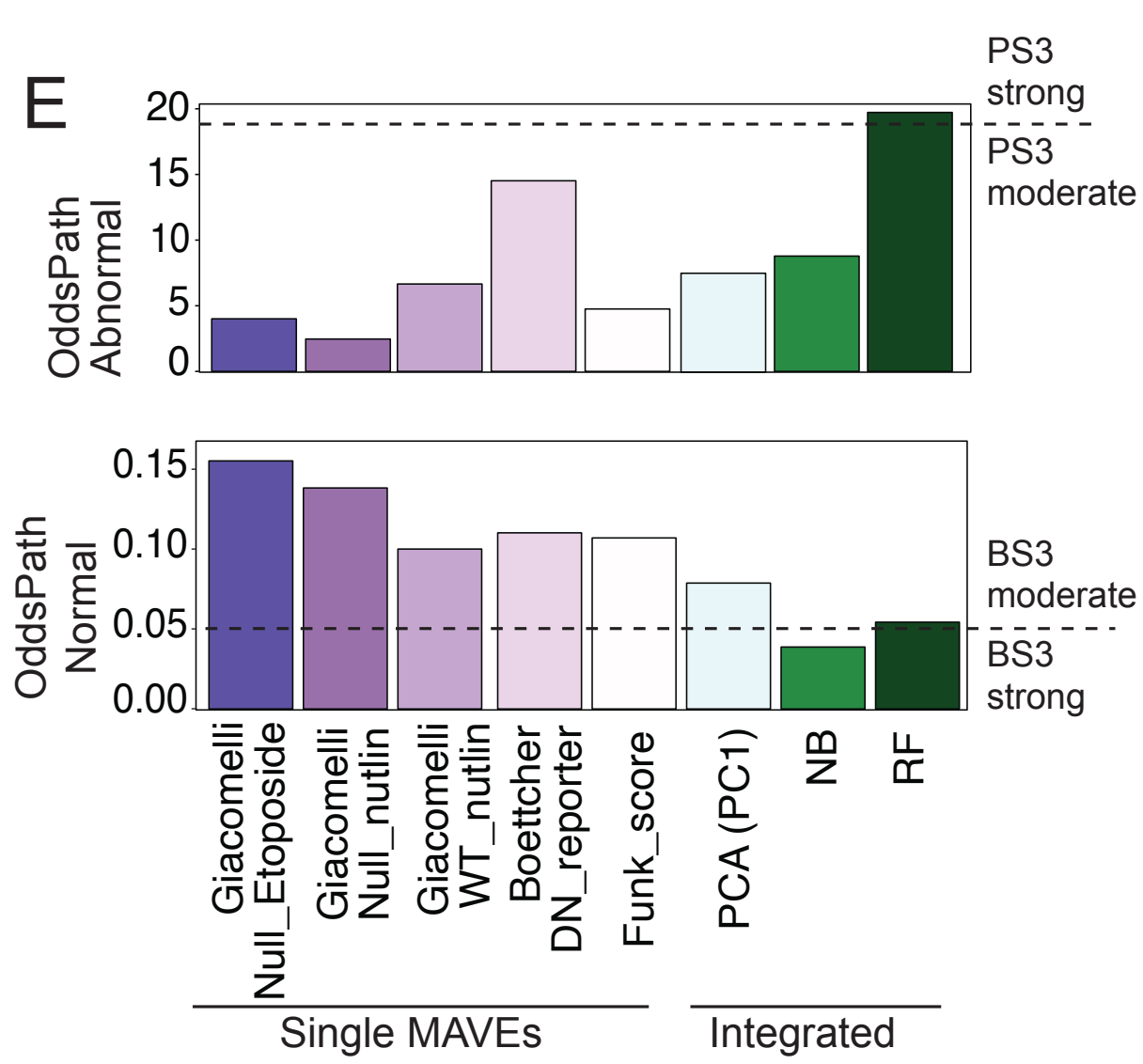

**Figure 3: Example of combining multiplexed functional data for *TP53* MAVEs. (A)** The number of variants of different classes present in the constituent assays is indicated; P/LP and B/LP variants correspond to the clinical truth set; PTC and Syn variants constitute the proxy truth set. "Ratio" refers to the fraction of clinical truth set combined with proxy truth set for the purpose of model training (e.g., 0.7/0.01 = 70% of clinical variants, 1% of proxy variants). (**B**) Principal component analysis of the scores in five constituent *TP53* datasets. The first two principal components are shown. Variants were clustered by class within the dataset for easy visualization. **(C)** Example K-means clustering output plotted to visualize candidate clusters (n=4) within principal component space. **(D)** Comparison of sensitivity, specificity, PPV and NPV for individual *TP53* MAVEs with integrated scores using different methodologies. **(E)** For each method, we calculated the OddsPath for variants classified as functionally abnormal **(top)** or normal **(bottom)** to determine what strength of evidence could be applied towards pathogenicity or benignity, respectively, compared to the best-performing single MAVE.

0.931 and specificity ranging from 0.629 to 0.944. The DN_reporter MAVE(42) performed best overall, achieving the highest specificity (0.944) and PPV (0.869) (**Figure 3D**). A classifier based on PC1 of the Principal Component Analysis did not improve specificity compared to the best performing individual MAVE, though it did subtly improve sensitivity. The naïve Bayes classifier improved specificity and achieved BS3_strong evidence strength for OddsPath normal (**Figure 3E**), with a 3-fold cross validation kappa value of 0.752. The random forest classifier, on the other hand, improved specificity and sensitivity, as well as led to significant improvement in OddsPath for both functionally abnormal and normal variants, achieving PS3_strong evidence strength (**Figure 3E;** 3-fold cross validation z-score = 0.098). Note that both supervised methods do have some degree of overfitting to training data in identifying benign variants as functionally normal, as shown by the specificity value of 1 vs. ~0.9 in the train/test split. These results suggest that while integration may not improve basic classification metrics, it can enhance clinical utility through stronger evidence assignments, though careful validation and potential model refinement are needed to address overfitting issues.

For *LDLR*, we used two recently published MAVEs (32). One MAVE measured steady-state protein abundance at the cell surface while the other measured gene activity (41, 42). We analyzed these datasets with a clinical truth set of 371 missense variants (15 B/LB and 356 P/LP) (**Figure 4A**). As this is a relatively unbalanced truth set, we included 20% of the available proxy assay control variants to aid in model training. 50% of missense variants were used for

A

| *LDLR* Truth Set | P/LP | B/LB | PTC | Syn | Ratio |
|---|---|---|---|---|---|
| | 356 | 15 | 774 | 685 | 0.5/0.2 |

B

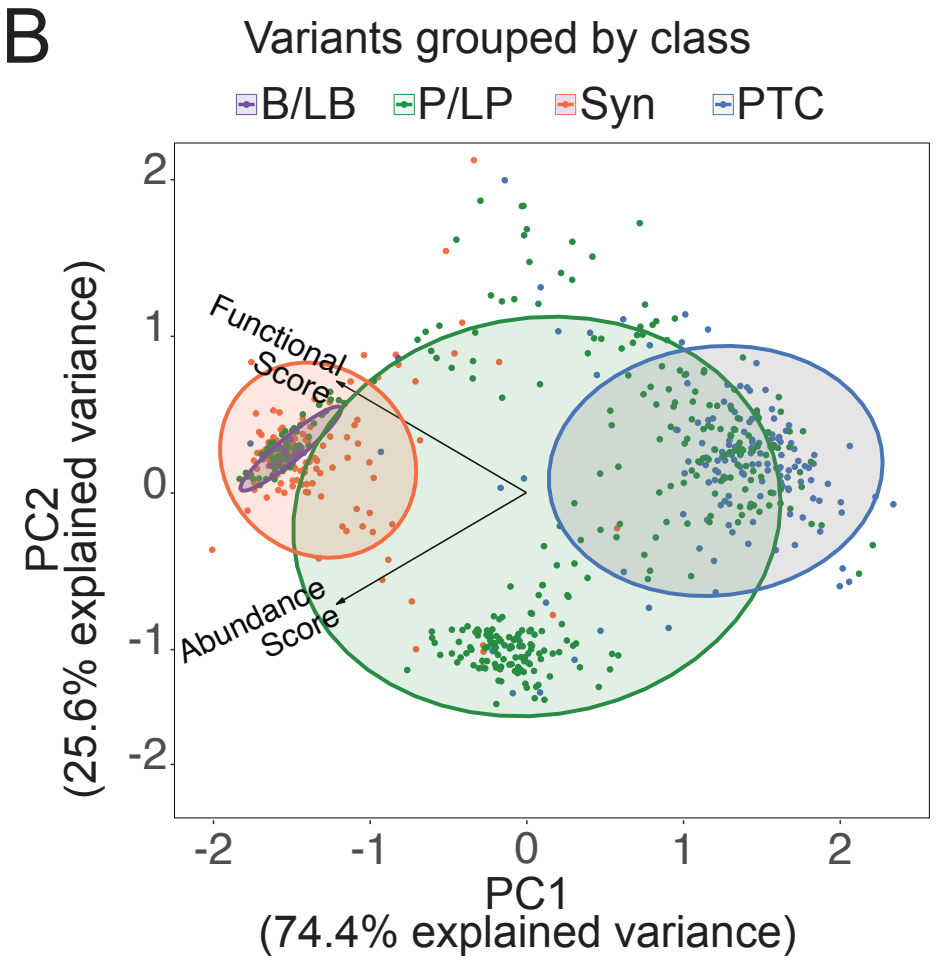

C

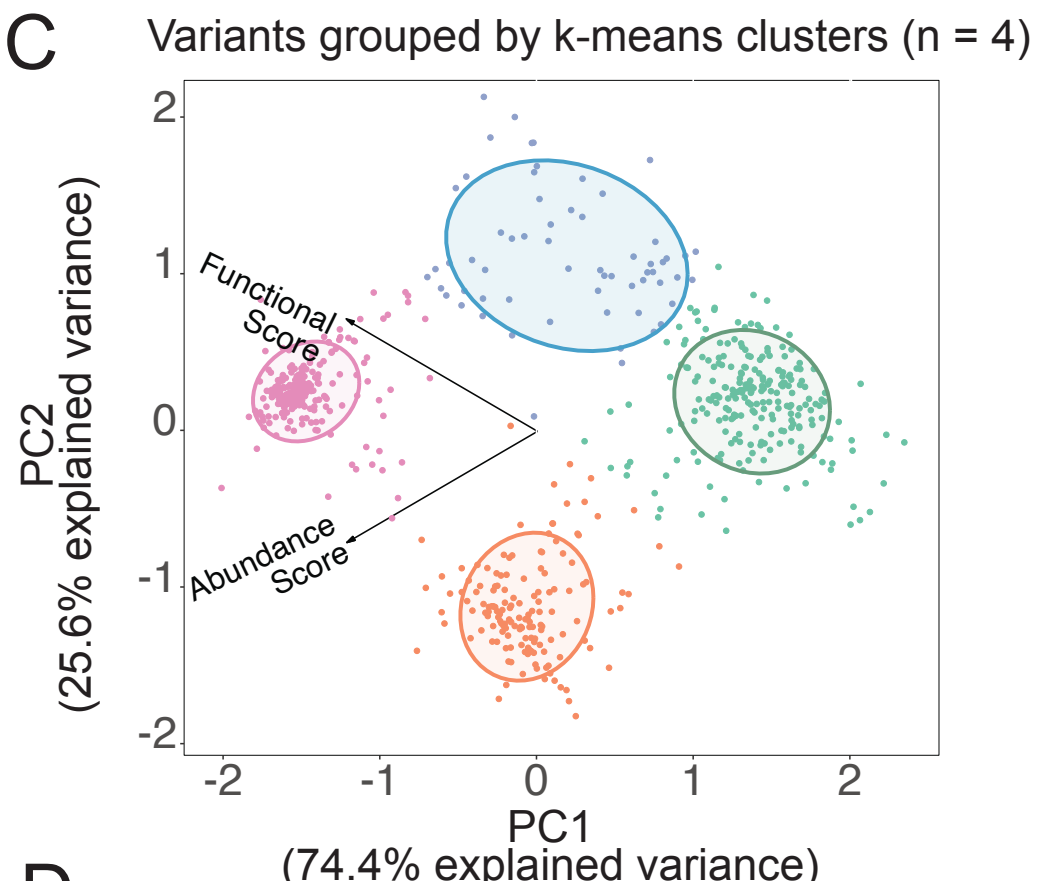

D

| *LDLR* | Sens. | Spec. | PPV | NPV | AUC |
|---|---|---|---|---|---|
| MAVE1: Abundance | 0.604 | 0.933 | 0.995 | 0.090 | 0.713 |
| MAVE2: Function | 0.787 | 1.000 | 1.000 | 0.165 | 0.871 |
| PCA | 0.795 | 1.000 | 1.000 | 0.170 | 0.898 |
| Naive Bayes (train) | 0.802 | 1.000 | 1.000 | 0.186 | 0.879 |
| Naive Bayes (test) | 0.788 | 1.000 | 1.000 | 0.156 | 0.931 |
| Random forest (train) | 0.989 | 1.000 | 1.000 | 0.800 | 0.997 |
| Random forest (test) | 0.844 | 1.000 | 1.000 | 0.200 | 0.911 |

E

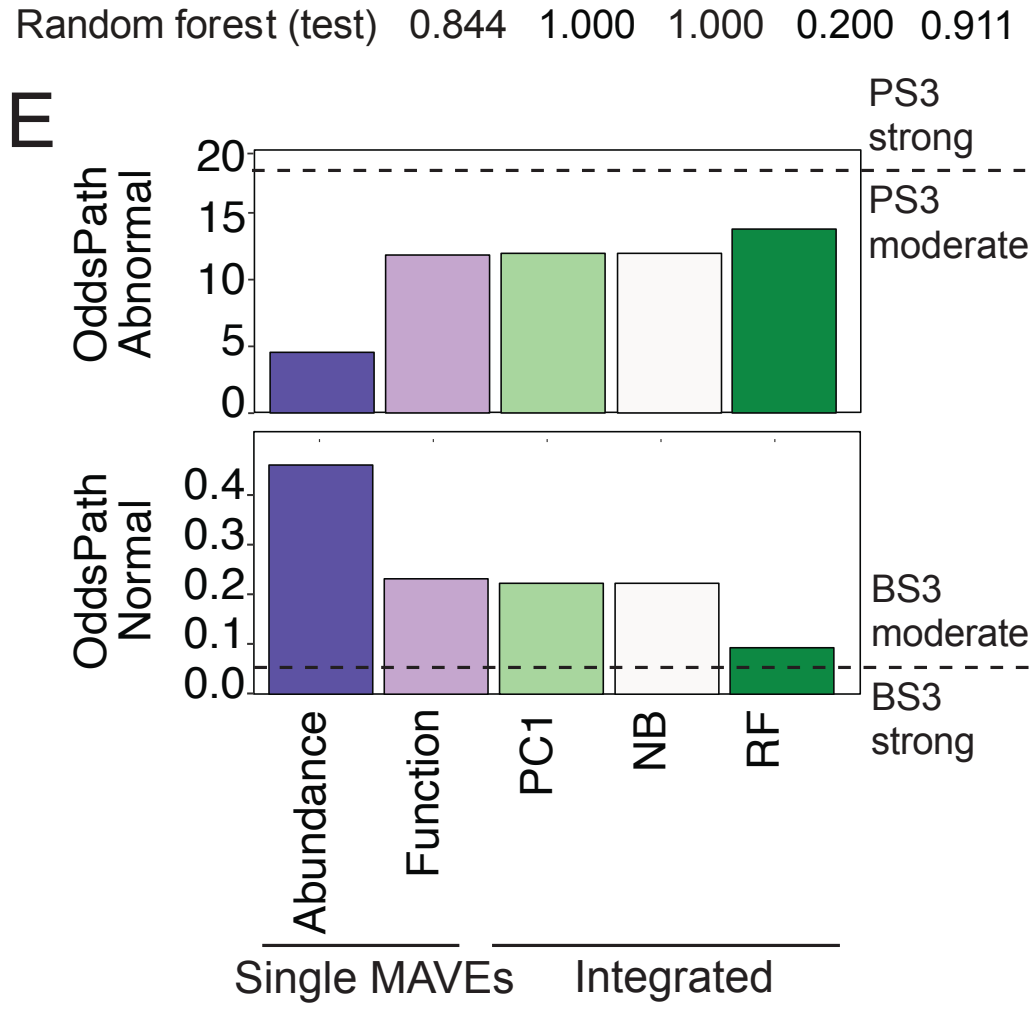

F

| *PTEN* Truth Set | P/LP | B/LB | PTC | Syn | Ratio |
|---|---|---|---|---|---|
| | 105 | 16 | 319 | 87 | 0.4/1.00 |

G

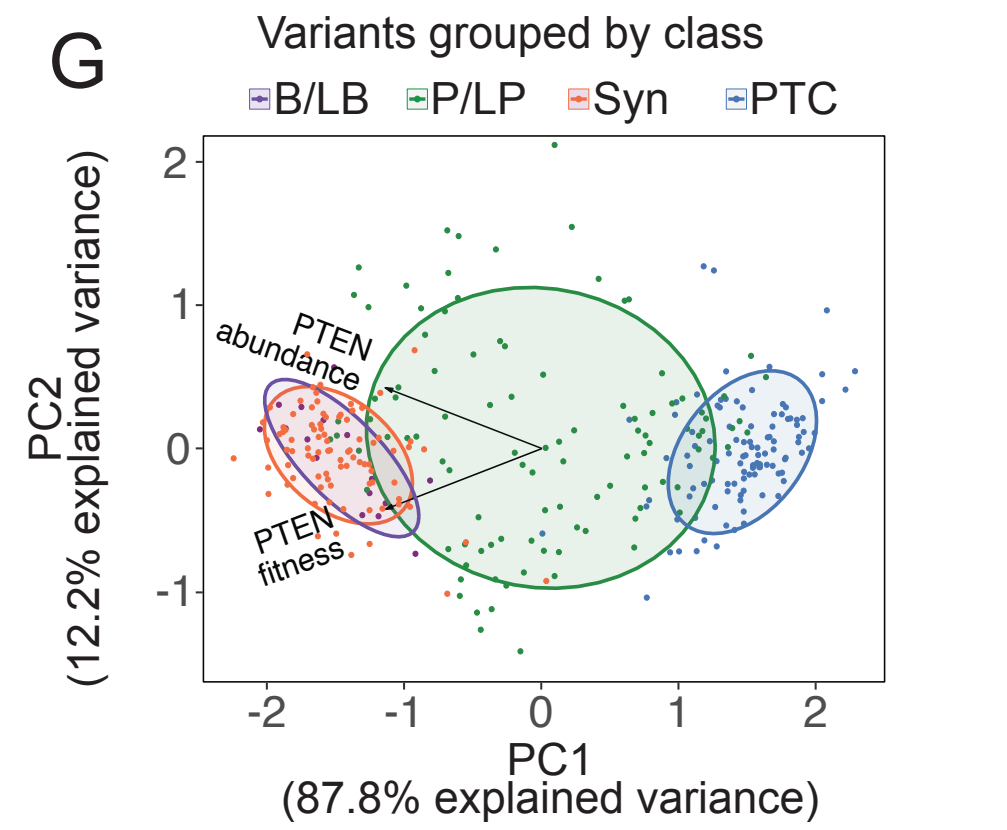

H

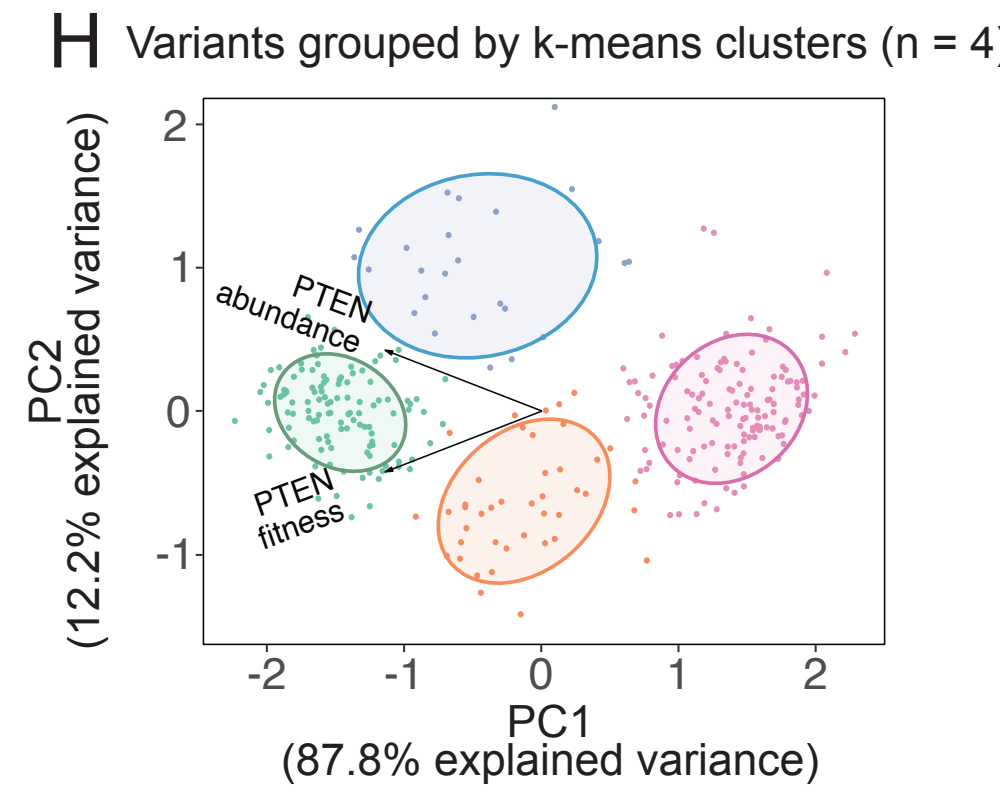

I

| *PTEN* | Sens. | Spec. | PPV | NPV | AUC |
|---|---|---|---|---|---|
| MAVE1: Abundance | 0.619 | 1.000 | 1.000 | 0.286 | 0.819 |
| MAVE2: Fitness | 0.790 | 1.000 | 1.000 | 0.421 | 0.943 |
| PCA | 0.819 | 1.000 | 1.000 | 0.457 | 0.952 |
| Naive Bayes (train) | 0.850 | 1.000 | 1.000 | 0.571 | 0.984 |
| Naive Bayes (test) | 0.892 | 1.000 | 1.000 | 0.533 | 0.952 |
| Random forest (train) | 0.975 | 1.000 | 1.000 | 0.889 | 1.000 |
| Random forest (test) | 0.815 | 1.000 | 1.000 | 0.400 | 0.933 |

J

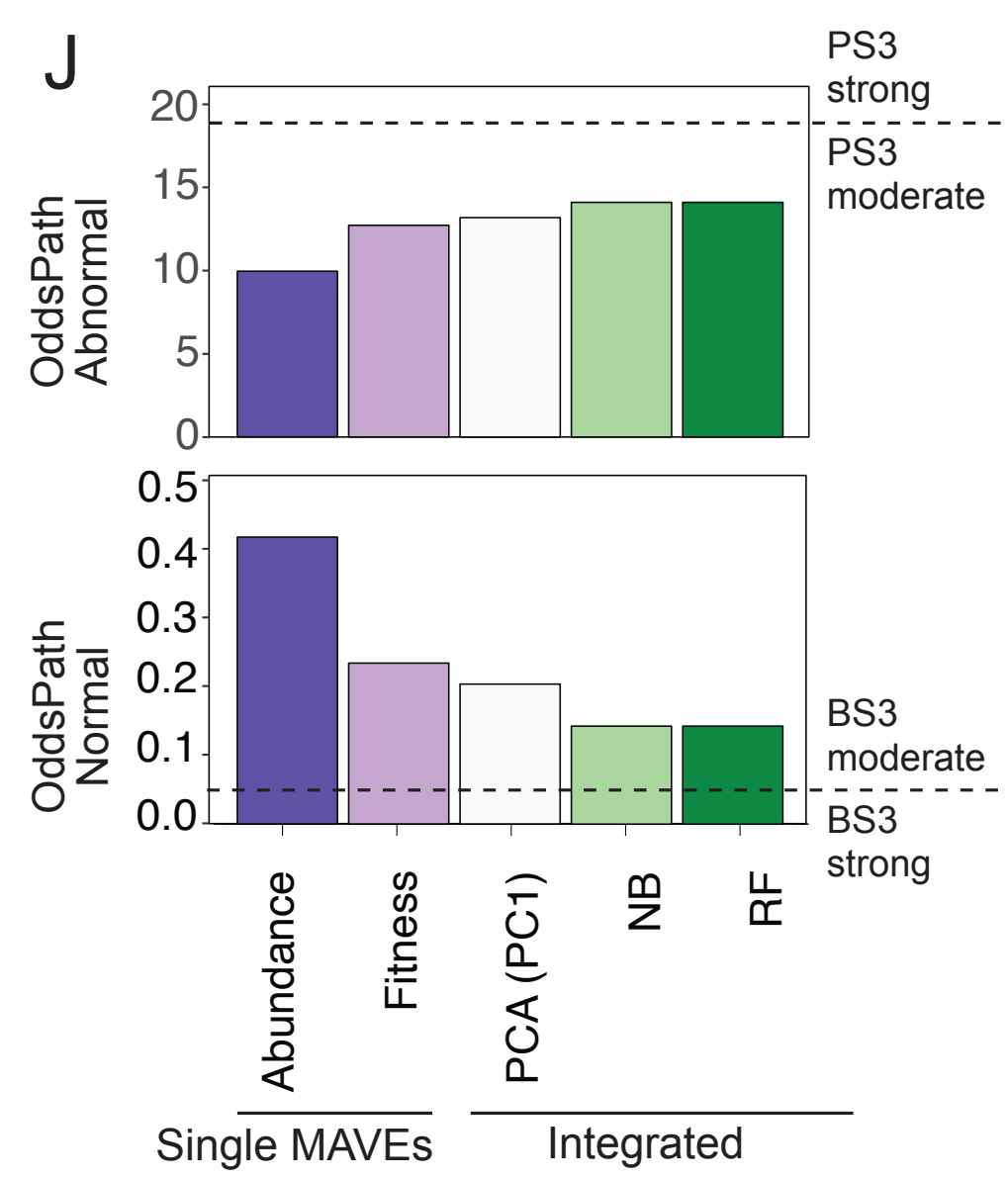

**Figure 4: Example of combining multiplexed functional data for *LDLR* and *PTEN* MAVEs.** **(A)** The number of variants of different classes present in the constituent assays is indicated; P/LP and B/LB variants correspond to the clinical truth set; PTC and Syn variants constitute the proxy truth set. "Ratio" indicates the proportion of clinical truth set/proxy truth set used in model training (e.g., 0.5/0.2 = 50% of clinical variants, 20% of proxy variants). (**B**) Principal component analysis of the scores in two constituent *LDLR* datasets. The first two principal components are shown. Variants were clustered by class within the dataset for easy visualization. **(C)** Example K-means clustering output plotted to visualize candidate clusters (n=4) within principal component space. **(D)** Comparison of sensitivity, specificity, PPV and NPV for individual *TP53* MAVEs with integrated scores using different methodologies. **(E)** For each method, we calculated the OddsPath for variants classified as functionally abnormal **(top)** or normal **(bottom)** to determine what strength of evidence could be applied towards pathogenicity or benignity, respectively, compared to the best-performing single MAVE. **(F-J)** Same plots as in (**A-E)** but for *PTEN*.

model training while the remaining 50% of variants were reserved as a test set.

Plotting the missense B/LB and P/LP variants in principal component space based on their scores in the two individual MAVEs, as well as based on K-means clusters (**Figure 4B-C**). reveal distinct clusters that distinguish B/LB and P/LP variants. Individual MAVEs showed variable performance, with sensitivity ranging from 0.604 to 0.787 and specificity ranging from 0.933 to 1. The activity MAVE (42) performed best overall, achieving the highest specificity (1.00) and sensitivity (0.787) ((**Figure 4D;** See **Tables S6-S10** for classifier outputs, threshold values, and comprehensive model metrics). A classifier based on PC1 of the principal component analysis slightly improved sensitivity compared to the best-performing individual MAVE due to three fewer false negative variants. The naïve Bayes classifier performed similarly to PC1, with similar sensitivity and specificity between training and test splits (**Figure 4E**). The random forest classifier, on the other hand, improved sensitivity without sacrificing specificity for the full truth set as well as each individual training and test split. For the full truth set, the random forest led to improvement in OddsPath for both functionally abnormal and normal variants (**Figure 4E**), although neither score crossed the threshold to count as “strong” evidence strength. It is important to note that there is some discrepancy in performance between the random forest on the training and test splits, suggesting mild overfitting (**Figure 4D**). We posit that the results presented herein argue for the potential of supervised learning to integrate the *LDLR* MAVEs. However, further work would be needed to optimize the random forest model to minimize overfitting.

For *PTEN*, we used two datasets that measured protein abundance and fitness effects of different variants (36, 37, 46). A clinical truth set of 38 B/LB variants and 216 P/LP variants was retrieved from ClinVar (accessed on 11 June 2025), with the caveat that functional data may have contributed to variant classification for some of these variants, potentially introducing bias and circularity to this analysis. The truth set variants were merged with the *PTEN* multiplexed functional datasets to obtain variants scored in both datasets (16 B/LB and 105 P/LP variants; **Figure 4F**). As this truth set is heavily biased to P/LP variants, we included all synonymous (presumed benign) and nonsense (presumed pathogenic) variants included in the MAVE as part of our proxy truth set, along with 40% of the clinical truth set (**Figure 4F**). Note that this proxy truth set is also biased towards presumed pathogenic variants (319 PTC; 87 Syn).

Principal component analysis and K-means clustering show very little overlap between presumed pathogenic variants (PTCs) and clinically validated P/LP variants, suggesting that clinical P/LP variants exhibit more nuanced functional differences than complete loss-of-function (i.e. PTC variants) (**Figure 4G-H**). In contrast, presumed benign variants (synonymous) cluster closely with clinically validated B/LB variants in PC space, which may be due to the small number of variants in both classes, as well as the potential use of functional data in ascertaining the benignity of the B/LB variants, i.e. circularity (**Figure 4G**).

Individual assays as well as the integrated models have relatively modest sensitivity but very high specificity, i.e. the pathogenic/abnormal calls are more reliable than the benign/normal calls. (**Figure 4I;** see **Tables S11-S15** for classifier outputs, threshold values, and comprehensive model metrics) This is also likely due to the small number of B/LB variants in the clinical truth set for *PTEN* as well as the relatively skewed distribution of nonsense and synonymous variants among assay controls (i.e. proxy truth set). PCA modestly improved sensitivity (0.819) while maintaining perfect specificity. Naive Bayes achieved reasonable training performance (sensitivity=0.850, NPV=0.571) that was maintained on test data (sensitivity=0.892, NPV=0.533). The random forest model exhibited some evidence of overfitting: sensitivity degraded from 1.000 to 0.908 and NPV from 1.000 to 0.571 in the test data compared to training data. Moreover, the integrated scores do not provide a higher PS3 strength of evidence for pathogenicity or benignity compared to individual MAVEs and therefore do not provide increased clinical utility (**Figure 4J**). The fundamental limitation here is likely the inadequate number of clinically validated B/LB variants, which prevents robust model training and calibration. Based on these results, we recommend revisiting *PTEN* data integration when additional clinically validated B/LB variants become available.

In summary, by evaluating various model metrics and the final OddsPath values for pathogenicity and benignity, we offer a framework to assess whether an integrated functional score generated by a specific model is reliable and has practical utility. Our examples show integration can have moderate (*TP53*) to mild (*LDLR* and *PTEN*) benefit, depending on the specific gene and the availability of robust truth sets for model training. We anticipate that the Shiny app we have developed will allow easy access to a set of standard integration methods to lower the barrier for similar exploratory analyses by end users.

**Alternative approaches in the literature**

Here, we have taken the approach where the input is a data matrix containing functional scores from independent assays, and the output is a single integrated score for each variant. This approach allows the calculation of an OddsPath to assign PS3/BS3 evidence codes across all variants at a particular calculated evidence strength. It is worth acknowledging that there are other frameworks for integrating multiple lines of functional evidence in variant classification, including gene-specific approaches and decision tree models developed for specific genes such as DNA mismatch repair genes, *TP53*, and *BRCA1* (68-71). Moreover, the OddsPath framework does have certain caveats. The OddsPath value represents the likelihood ratio for all variants above or below a set threshold for functionality, rather than for a specific variant at a particular score. Therefore, it is prone to overestimating confidence near the threshold and underestimating confidence far from the threshold. An example of an alternative approach is the log-likelihood ratio method used in van Loggerenberg et al (72), which may be particularly useful when variant scores show a wide dynamic range. The log-likelihood ratio approach is more granular; each variant is ascribed a PS3/BS3 evidence code at a different evidence strength. Using this approach on the *HMBS* gene allowed assignment of PS3/BS3 ACMG/AMP codes at either moderate or supporting evidence strength based on thresholds derived from the log-likelihood ratios (72).

## Discussion

*Limitations and future directions for integrating multiplexed functional data*

The methodologies we have applied to explore integrating multiplexed functional datasets and the shiny application developed for this purpose could be improved in several ways: **(1)** Support for analyzing variants that are only sampled by a subset of assays needs to be enabled for methods that permit missing values, such as a custom implementation of naïve bayes classifier that ignores missing attributes; **(2)** The ability to integrate normal/abnormal calls from low-throughput "gold standard" functional assays used by ClinGen variant curation expert panels (VCEPs) and any assays that produce categorical labels rather than continuously distributed scores; **(3)** More nuanced models designed for combining specific types of data, such as time course data (73, 74) or imaging data; **(4)** Implementing a framework for quantitative integration of standard error metrics from functional assays into variant categorization evidence strength. By making the code for the shiny application available on GitHub (an open-source platform for collaborating on software development), we hope that community-driven updates will add functionality to this tool over time.

*Dissemination of integrated multiplexed functional data and unification of clinical application*

Clinical application of an integrated score representative of multiple MAVEs requires the deposition of results into the public domain such as MaveDB (75). It is imperative that the integrated functional scores be clearly signposted as being derived from a MAVE combination to avoid double counting of data from both combined datasets and their constituent studies (a functionality built into MaveDB (75)), and labelled with respect to the relevant gene-disease dyad for clinical use to guard against inappropriate application. Similarly, combination methodologies that incorporate additional evidence sources, such as *in silico* variant effect predictions or allele frequencies, must be clearly highlighted to ensure evidence codes relating to contributing evidence types are not applied. Where possible, the evidence strengths that can be applied towards pathogenicity (PS3) and benignity (BS3) for the combined datasets under the Brnich et al. framework (20) should also be highlighted to steer appropriate clinical application. Consistency in the integration of multiplexed functional data would further benefit from expert review and incorporation into clinical guidelines, such as by a relevant ClinGen

VCEP. This may include iterative evaluation of MAVE performance against up-to-date truth sets as the number of classified variants available to comprise these truth sets increases.

*Improved classification of reduced penetrance variants and variable expressivity*

Most MAVEs produce a quantitative measurement of variant effect. However, the OddsPath framework converts this to a categorical evidence strength for use in clinical variant classification. The extent to which MAVE scores reflect quantitative differences in variant function, and whether this is relevant for human phenotypes in health and disease is currently being explored (22, 76, 77). Classification of variants falling in the intermediate range of MAVE scoring remains particularly challenging owing to difficulty in distinguishing truly intermediate readout from technical noise. Combined MAVE approaches offer a powerful way to resolve the effects of such variants: combining may mitigate the effect of experimental noise and allow assignment of evidence towards pathogenicity or benignity under a fully penetrant model. Alternatively, consistent observations of intermediate variant effect between assays or increasing variant effect along a time course assay (as may be the case with hypomorphs) may indicate the incomplete penetrance and/or variable expressivity of that variant (10, 12). As clinical guidance around classification and reporting of such variants continues to develop (78), integration of multiplexed functional datasets may improve certainty in assignment of hypomorphic functional impact.

*Insights into the genetic basis of disease*

Outside the clinical context, MAVE combination may yield further insights into biological mechanisms of disease: as discussed above, unsupervised learning approaches may allow clustering of variants with shared characteristics that may not be apparent when considered at the individual assay level. Variant clusters that do not represent known relationships between variant function under different assay models may allow elucidation of novel biological mechanisms (79). When MAVEs underpinned by the same assay methodology but using different biological models (e.g. different cell types) are combined, investigation of specific variant clusters may allow insight into the underlying biological differences between the models and allow hypothesis generation for future research and possible refinement of clinical interpretation. Similarly, appropriate integration of multiplexed functional data generated in models representing different time points (e.g. stages of development) (12) may allow a better understanding of the temporal dynamics of variant effect (10), which may be particularly insightful in the investigation of many developmental disorders, which take place against a rapidly changing genomic and transcriptomic program (9, 10, 12).

**Conclusion**

The rapid increase in variants of uncertain significance (VUS) classifications in recent years has highlighted the urgent need for efficient, large-scale methods to resolve variant classification. MAVEs have emerged as powerful tools to address this challenge, allowing for the functional assessment of hundreds to thousands of variants in a single experiment. As the number of published MAVEs for individual genes continues to grow, the potential to combine data from multiple assays presents a promising avenue for improving variant classification accuracy. Moreover, the forthcoming v4.0 update to the ACMG/AMP guidelines is expected to be based on a more continuous evidence point-based system (80), with the existing boundaries between categorical evidence strengths being replaced by integer LLR value thresholds. Any improvements in applicable evidence through a combination of multiplex data, however modest, are more likely to cross these integer thresholds and improve applicable evidence strength than to cross the existing categorical thresholds and their LLR equivalents. By providing a framework for when and how to combine multiplexed functional data and an easy-to-use web tool to test different methods of integrating multiplexed data, we aim to accelerate progress in this field and ultimately improve patient care through more accurate and efficient genetic diagnoses.

**Acknowledgments**

The authors would like to thank Fritz Roth, Kresten Lindorff-Larsen, Ben Livesey, Joseph Marsh, and Ashley Marsh, as well as all members of the Atlas of Variant Effects Clinical Variant Interpretation working group, for providing helpful suggestions and feedback on the manuscript. We thank Lara Muffley for assistance with project management. S.J. was supported by NIH R35 GM133433. L.M.S. was supported by R01HG013025 and UM1HG011969. A.B.S. was supported by an NHMRC Investigator Fellowship (APP177524). C.T. and C.F.R. are supported by CRUK Programme Award CG-MAVE (EDDPGM Nov22/100004). S.F. was supported by UM1HG011969 and RM1HG010461. M.D. was supported by U01HG011758, UM1HG011969, RM1HG010461, OT2OD002751, and CPRITRP210027. A.E.M. was supported R01HG013025,

an Early Career Award from the Alex's Lemonade Stand for Childhood Cancer and RUNX1 foundation (21-25037), and the Brotman Baty Institute Catalytic Collaborations Grant (CC28).

**Data and code availability.** The web Shiny application is available at https://jagannathan-lab.shinyapps.io/AVE_shinyApp_Jun2025/.The Github repository with Shiny application code, example csv input files, and detailed instructions are available at https://github.com/jagannathan-lab/2025-calhoun_et_al.

**Declaration of Interests.** The authors declare that they have no competing interests.

**Declaration of generative AI and AI-assisted technologies in the writing process.** During the preparation of this work, the authors used Claude.ai in order to improve readability, flow and language. After using this tool, the authors reviewed and edited the content as needed and take full responsibility for the content of the publication.